\documentclass[11pt,onecolumn,amssymb,nofootinbib]{revtex4}
\usepackage{amsmath, amsthm, amscd, amssymb}
\usepackage{bm}
\usepackage{bbm}
\usepackage{slashed}

\begin{document}

\title{\bf Soft Pions and More}

\author{Stephen L. Adler}
\email{adler@ias.edu} \affiliation{Institute for Advanced Study,
Einstein Drive, Princeton, NJ 08540, USA.}

\begin{abstract}
I review the role that soft pion theorems, current algebras, sum rules, and anomalies played in the foundation of the standard model.
\end{abstract}

\maketitle

\section{Outline of Sections}

My talk is divided into sections as follows, with the focus in each section on a few key papers and formulas. For more details and citations, beyond
those given in the individual sections, see the ``Commentaries'' in my book ``Adventures in Theoretical Physics'', Vol. 37 in the World Scientific Series in
20th Century Physics, Chapters 2 and 3.  I wish to thank Bryan Lynn and Glenn Starkman, co-chairs of the organizing committee of the ``The Standard Model at 50 Years'' Symposium, held June 1-4 at Case Western Reserve University, for inviting me to talk. Their recent work has incorporated a  modern incarnation of the ``Adler zeros'' discussed below.
I am also grateful to Harsh Mathur for overseeing the Symposium proceedings and for supplying a transcript which facilitated preparation of this written version, and to
my former collaborator Bob Brown for chairing the session at which I spoke.

\bigskip
\begin{itemize}
\item{Preliminaries}
\item{Soft Pions}
\item{Current Algebras}
\item{Sum Rules}
\item{Further Soft Pion Applications}
\item{Deep Inelastic Scattering Sum Rules}
\item{Anomalies and PCAC}
\item{Historical Significance}
\item{Updates and Revivals}
\end{itemize}
\section{Preliminaries}

Let us begin with a review of Noether's Theorem.  Consider a Lagrangian density ${\cal L}$ that is a function of fields $\Phi_{\ell}$ and their space-time derivatives
$\partial_{\lambda}\Phi_{\ell}$,
\begin{equation}\label{lagrangian}
{\cal L}(x)= {\cal L}[\Phi_{\ell},\partial_{\lambda}\Phi_{\ell}].
\end{equation}
Under a transformation of the fields
\begin{equation}\label{transform}
\Phi_j(x) \to \Phi_j(x) + \Lambda(x)G_j[\Phi_{\ell}]
\end{equation}
the first order change in the Lagrangian density is, by applying the chain rule,
\begin{equation}\label{firstorder}
\delta {\cal L}=\frac{\delta {\cal L}}{\delta \Lambda} \Lambda + \frac{\delta {\cal L}}{\delta(\partial_{\alpha}\Lambda)}\partial_{\alpha}\Lambda~~~.
\end{equation}
A simple calculation using the Euler-Lagrange equations now shows that if we define a  current $J^{\alpha}$ by
\begin{equation}\label{current}
J^{\alpha}=-\frac{\delta {\cal L}}{\delta(\partial_{\alpha}\Lambda)}=-\sum_j\frac{\delta {\cal L}}{\delta(\partial_\alpha \Phi_j)} G_j~~~,
\end{equation}
then the divergence of the current is given by
\begin{equation}\label{divj}
\partial_\alpha J^\alpha=-\frac{\delta {\cal L}}{\delta \Lambda}~~~,
\end{equation}
which is the first Noether theorem.
Thus if ${\cal L}$ is  invariant under the transformation of Eq. \eqref{transform} for constant $\Lambda$, then ${\delta \Lambda}=0$ and the current is {\it conserved}, while if  ${\delta \Lambda}\neq 0$ but  small, the current is said to be {\it partially conserved}.

Let us now apply this to the Dirac Lagrangian,
\begin{equation}\label{dirac}
{\cal L}=\bar\psi \gamma^\mu \partial_\mu \psi + i m \bar \psi  \psi, ~~~~ \bar \psi=\psi^\dagger \gamma^0~~~.
\end{equation}
Under the phase transformation
\begin{equation}\label{phase}
\psi \to e^{i\theta}\psi
\end{equation}
${\cal L}$ is invariant, which implies that there is a  {\it conserved vector current},
\begin{equation}\label{vector}
V^\mu=\bar \psi \gamma^\mu  \psi~,~~~~\partial_\mu  V^\mu=0~~~.
\end{equation}
On the other hand, under the chiral phase transformation
\begin{equation}\label{chiralphase}
\psi \to e^{i\theta\gamma_5}\psi~~~,
\end{equation}
since $[\gamma_5,\gamma^0\gamma^\mu]=0$ but $[\gamma_5,\gamma^0]\neq 0$, the Dirac Lagrangian density is invariant only when $m=0$.
So there is a {\it partially conserved axial-vector current}
\begin{equation}\label{axial}
A^\mu=\bar \psi \gamma^\mu \gamma_5 \psi~,~~~~\partial_\mu  A^\mu=im\bar \psi \gamma_5 \psi~~~.
\end{equation}

\section{Soft Pions}

\subsection{Goldberger-Treiman Relation}
{\bf Featured paper:}  ``Decay of the Pi Meson'', by M. L. Goldberger and S. B. Treiman,  Palmer Physical Laboratory, Princeton University,
Phys. Rev. {\bf 110}, 1178 (1958).

This is the paper that, from a slightly obscure dispersion relation analysis, derived a remarkable formula now known as the {\it Goldberger-Treiman relation}.
(Incidentally, Sam Treiman later on was my thesis advisor. I knew both him and Goldberger well when I was a graduate student at Princeton.)
The Goldberger-Treiman relation states that\footnote{Equation \eqref{gtrel} uses the Particle Data Group definition of $f_\pi$, which is the $f_\pi$ of the
Adler-Dashen book {\it Current Algebras} divided by $M_\pi^2$.}
\begin{equation}\label{gtrel}
f_\pi=\frac{\surd 2  M_N g_A}{g_r}~~~,
\end{equation}
where
\begin{align}\label{defs}
f_\pi=&{\rm charged~pion~decay~constant}~~~\cr
g_A=&{\rm nucleon~axial-vector~coupling~constant}~~~\cr
g_r=&{\rm pion-nucleon~coupling~constant}~~~\cr
M_N=&{\rm nucleon~mass}~~~.\cr
\end{align}
This relation works amazingly well.  It was good to around 6\% when first derived, and now with current numbers is good to 2.3\%.

\subsection{Reinterpretation}

Two years later, in 1960, the Goldberger-Treiman relation was reinterpreted as a partially-conserved or pion pole dominated axial-vector current,
by many authors (Nambu; Bernstein, Fubini, Gell-Mann and Thirring; Gell-Man and L\'evy; Bernstein, Gell-Mann and Michel; Chou),
\begin{equation}\label{partial}
\partial_\lambda A^\lambda_{1+i2}=\partial_\lambda {\cal F}^{5\lambda}_{1+i2}=c\phi^\dagger_{\pi^+}=c\phi_{\pi^-}~,~~~c=\frac{\surd2 M_N g_A}{g_r(0)}\simeq \frac{\surd 2 M_N g_A}{g_r}\simeq f_\pi~~~,
\end{equation}
with $\phi^\dagger_{\pi^+}$ the creation operator for a positively-charged pion, and with ${\cal F}^{5\lambda}_{1+i2}$ the Gell-Mann notation for the axial-vector current that we shall use from now on.

\subsection{Early Applications}

{\bf Featured Paper:}  ``Chirality Conservation and Soft Pion Production'', by Y. Nambu and D. Luri\'e,  The Enrico Fermi Institute for Nuclear Studies and the Department of Physics, University of Chicago, Phys. Rev. {\bf 125}, 1429 (1962).

This paper considered the limit of an exactly conserved axial-vector current and defined a chirality $\chi$ as the nucleon plus pion axial-vector charge (in their notation, $\chi=-i\int d^3x j_4$).  They then showed that exact conservation of this chirality implies that in any reaction,
\begin{equation}\label{nambu}
\langle \alpha^{\rm in}|\chi|\alpha^{\rm in}\rangle= \langle \alpha^{\rm out}|\chi|\alpha^{\rm out}\rangle~~~,
\end{equation}
and  applied this to relate the amplitude for the soft pion production reaction $\pi+N \to \pi+N+\pi({\rm soft})$ to the amplitude for pion-nucleon scattering
$\pi+N \to \pi+N$. Their relation worked to within about a factor of three but not better, and that was probably one of the reasons why this didn't really catch on at the time; there wasn't follow-up work. The other was that there was an  an undetermined renormalization constant relating their chirality to the axial charge, which was not pinned down in their analysis. I later showed that this is just $g_A$. So there are some pieces in the Nambu-Luri\`e paper that were a bit obscure, and  this paper, which was a predecessor of all soft pion theorems, kind of languished.

{\bf Featured Paper:}  ``Tests of the Conserved Vector Current and Partially Conserved Axial-Vector Current Hypotheses in High-Energy Neutrino Reactions'', by
Stephen L. Adler, Princeton University, Phys. Rev. {\bf 135}, B963 (1964).

This paper relates the differential cross section $d\sigma/d\Omega(\nu +\alpha \to \ell+\beta)$ for neutrino scattering with a forward lepton $\ell$ to the cross section $\sigma(\pi+\alpha \to \beta)$,
that is, $\nu \bar \ell$ at $q^2\simeq 0$ is like a $\pi$ by conservation of the vector current and partial conservation of the axial-vector current, when the lepton mass can be neglected.  This is a hard experiment to do, and so this relation was only tested decades after I wrote the paper and after much other soft pion work.  At the time I wrote this paper, CVC was already a popular
acronym for conserved vector current, and so in writing  this paper I introduced the term PCAC because I thought the partially conserved axial-vector current should be popularized by giving it another catchy acronym.

\subsection{Soft Pion Theorems as a Precision Tool}

{\bf Featured Paper:}  ``Consistency Conditions on the Strong Interactions Implied by a Partially Conserved Axial-Vector Current'', by Stephen L. Adler, Palmer Physical
Laboratory, Princeton University, and Bell Telephone Laboratories, Murray Hill, Phys. Rev. 137, B1022 (1964).

When I started looking at PCAC Sam Treiman said that there was only one number, the Goldberger-Treiman relation, but soon there would be more.
The first indication beyond the Goldberger-Treiman relation that soft pions can be a precision tool, was a paper I wrote, partly  while I was at Princeton, but written up with Phil Anderson's permission  when I was working at Bell Labs during the summer after I got my PhD. This paper came out of my thesis study of weak pion production; by manipulating some of  the amplitudes I found that PCAC implies the following relation,
\begin{align}\label{consistency}
A^{\pi N (+)}(\nu=&\nu_B=0)\simeq g_r^2/M_N~~~~,\cr
A^{\pi N (+)}=&{\rm symmetric~isospin~pion-nucleon~scattering~amplitude} ~~~.\cr
\end{align}
By a dispersion relation analysis using the measured pion-nucleon scattering phase shifts, I showed that this is good to about ten percent. The reason this fits in the soft pion  framework is that here you have $\pi+N \to N+ \pi({\rm soft})$, which gets related to $\pi +N \to N$, which is just the pion-nucleon coupling constant. So this is the simplest case of a soft pion theorem where you take one soft pion out and relate it to the reaction without the soft pion. I also showed in the same paper that if you look at pion-pion scattering, again with one  pion off shell, the corresponding answer is zero,
\begin{equation}\label{pipi}
A^{\pi\pi}(s=t=u=-M_{\pi}^2,k^2={\rm fourth~pion~mass}=0)=0.
\end{equation}
This is the origin of the so-called Adler zeros.

{\bf Featured Paper:} ``Consistency Conditions on the Strong Interactions Implied by a Partially Conserved Axial-Vector Current. II'', by Stephen L. Adler, Lyman Laboratory
of Physics, Harvard University, Phys. Rev. {\bf 139}, B1638 (1965).

In this second paper, I worked out the general external line insertion rules for  weak pion production.  Consider $A \to B + \pi({\rm soft})$, which can be related to
$A \to B$   without the soft pion. When there are external nucleon lines or in general external baryon lines - lambda, sigma, whatever - there's a non-zero insertion. But for pion lines, just because the three pion vertex is zero by the pseudo-scalar nature of the pion, there's a zero insertion, and that's the origin of the Adler zeros.

\section{Current Algebras}

{\bf Featured Paper:}  ``The Symmetry Group of Vector and Axial-Vector Currents'', by Murray Gell-Mann, California Institute of Technology, Physics {\bf 1}, 63 (1964).

So far I've just been considering PCAC by itself, now we come to current algebra.
There is a famous paper by Gell-Mann in the first volume of Physics, which was a journal that Phil Anderson started and which ran for a few volumes.

In Gell-Mann's paper, he defined a vector charge $F_i(t)$ as the spatial integral of the vector current fourth component $F_i(t)=-i\int d^3x {\cal F}_{i4}$, and an axial-vector charge $F_i^5(t)$ as an integral of the axial-vector current fourth component $F_i^5(t)=-i\int d^3x {\cal F}_{i4}^5$. The first two relations in his paper,
\begin{equation}\label{octet}
[F_i(t),F_j(t)]=if_{ijk}F_k(t)~~~~,~~~~[F_i(t),F_j^5(t)]=if_{ijk}F_k^5(t)
\end{equation}
with $f_{ijk}$ the $SU(3)$ structure constants, are just a statement that $F_i$ and $F_i^5$ are $SU(3)$ octets.
But then he introduced something revolutionary, by stating that  the commutation relations of the axial-vector charges close the algebraic system by giving  a vector charge, \begin{equation}\label{closure}
[F_i^5(t),F_j^5(t)]=if_{ijk}F_k(t)~~~.
\end{equation}
He arrived at this by abstracting from the simplest quark model, using the fact that $\gamma_5^2=1$.  This insight has had very important consequences.

{\bf Featured Paper:}  ``Renormalization Effects for Partially Conserved Currents'', by S. Fubini, Istituto de Fisica del l'Universit\`a-Torino and CERN, and
G. Furlan, CERN, Physics {\bf 1}, 229 (1965).

In the same initial volume of Physics, an important tool for exploiting current algegras was given in the paper by Fubini and Furlan.  What they did was to take the isospin raising vector charge, commuted with the isospin lowering vector charge, to give the isospin neutral vector charge. By CVC this latter has a matrix element of one, so you know what it is. And then they added a novel ingredient:  Instead of taking this matrix element at rest, they took it in the infinite momentum frame,
\begin{equation}\label{inf}
\langle N(\vec p\,)|[F_{1+i2},F_{1-i2}]|N(\vec p\,)\rangle~~~~,~~~{\rm limit}~~ p_z \to \infty~~~.
\end{equation}
Going to the infinite momentum frame in the vector-vector commutator case that they considered gives a nice kinematic simplification, but as we shall see, in the
axial-vector commutator case is absolutely crucial. Before going on to this, let me mention that in this same volume of Physics, there's also the famous Bell paper on the Bell inequalities, so it's a really classic volume.

\section{Sum Rules}

{\bf Featured Paper:}   ``Calculation of the Axial-Vector Coupling Constant Renormalization in $\beta$ Decay'', by Stephen L. Adler, Lyman Laboratory of Physics,
Harvard University, Phys. Rev. Lett. {\bf 14}, 1051 (1965).

{\bf Featured Paper:}  ``Renormalization of the Weak Axial-Vector Coupling Constant'', by William I. Weisberger, Stanford Linear Accelerator Center, Phys. Rev.
Lett. {\bf 14}, 1047 (1964).

The natural thing, knowing both the Gell-Mann current algebra and the Fubini-Furlan idea of the infinite momentum frame, is to apply them to the commutator of axial-vector charges, which is what was done by me and by Bill Weisberger.
You take the same commutator as in the Fubini-Furlan paper but with the vector isospin raising and lowering charges replaced by the corresponding axial-vector
charges.  This commutator, by Gell-Mann's algebra,  still closes into the neutral vector current charge, which has  a matrix element of one. Taking the $p_z$ to infinity limit is now crucially important because the axial-vector charge matrix element between nucleons at rest is zero by parity, but it's not zero in the infinite momentum frame, so you get something non-trivial. When you introduce a time derivative upstairs and a compensating denominator downstairs and use PCAC, you end up with the following remarkable sum rule,
\begin{equation}\label{aw}
g_A^2=1+f_\pi^2\, \frac{2}{\pi} \int_{M_N+M_\pi}^\infty \frac{WdW}{W^2-M_N^2}[\sigma^{\pi^+p}(W)-\sigma^{\pi^-p}(W)]~~~,
\end{equation}
with the integral from threshold to infinity on the right hand side involving the difference of $\pi^{\pm}p$ scattering cross sections at energy $W$.  This relation gives $g_A=1.24$; experiment now gives $g_A=1.272$, so we see we're now beginning to get good precision from PCAC.  When I showed this result to
Sam Treiman he said, well, now there's a third number,  with the consistency condition discussed above being the second number.  At this point other people  began to get interested.

{\bf Featured Paper:} ``Sum Rules for the Axial-Vector Coupling-Constant Renormalization in $\beta$ Decay'', by Stephen L. Adler, Lyman Laboratory of Physics, Harvard University, Phys. Rev. {\bf 140}, B736 (1965).

In 1965 when you wrote a Physical Review Letter,  you were supposed to write a  longer follow-up paper in Physical Review.    In my follow-up paper on
the $g_A$ sum rule, I also derived a $\pi \pi$ sum rule by sandwiching the same axial charge commutator between pion states in the infinite momentum frame,
\begin{equation}\label{pipi1}
1=f_{\pi}^2\, \frac{1}{\pi} \int_{2 M_\pi}^\infty  \frac{W dW}{W^2-M_\pi^2}[\sigma^{\pi^-\pi^+}(W)-\sigma^{\pi^+\pi^+}(W)]~~~,
\end{equation}
and showed that one needed a large isospin zero $S$-wave $\pi \pi$ cross section at low energy to saturate this sum rule.   This conclusion was basically qualitative, because
one did not have $\pi \pi$ scattering cross sections or phase shifts then.  In this paper I also derived a sum rule for forward lepton $(q^2\simeq 0)$ inclusive neutrino-nucleon scattering, by combining the $g_A$ sum rule which involves pion-nucleon scattering cross sections,  with the forward lepton theorem, which we recall says that forward lepton neutrino-nucleon scattering looks like pion-nucleon scattering.

\section{Further Soft Pion Applications}

 A flood of papers followed publication of the Adler-Weisberger sum rule, because there were now ``three numbers'':  Evidently  you could  get precision results from the combined ideas of PCAC and current algebra. Just to mention a few of the more prominent ones:
\begin{itemize}
\item{Callan and Treiman gave a relation between $K\to \pi \ell \nu$ and $K \to \ell \nu$.}
\item{Weinberg analyzed $K \to 2\pi \ell \nu$  and showed the importance of pion pole terms.   A pion pole term also enters into the calculation of $\pi N \to 2 \pi N$,
which is one of the reasons why Nambu and Luri\'e didn't get a good answer in their comparison with experiment.}
\item{ Weinberg also gave a discussion of $\pi\pi$ scattering lengths and multiple pion production.}
\end{itemize}

 The above papers are reprinted in the book I put together with Roger Dashen, {\it Current Algebras and Applications to Particle Physics}, Benjamin, 1968. There is also a nice discussion of soft pion theorems in Chapter Two of Sidney Coleman's collected lectures, {\it Aspects of Symmetry}, Cambridge University Press, 1985.

Up to this point I had never really heard of the idea of soft pion theorems as being an aspect of Nambu-Goldstone bosons.
That, to my cognizance, first appeared in a paper by Dashen developing chiral $SU(3)\times SU(3)$  as a strong interaction symmetry, and another paper on the same idea by Gell-Mann, Oakes and Renner.   These are both important papers generalizing soft pion theorems to the full $SU(3)$ octet of pseudoscalars,  and they emphasize the distinction between symmetry breaking of the axial-vector charge by parity doubling, as opposed to symmetry breaking by the appearance of a massless Goldstone boson, or an almost massless ``pseudo-Goldstone'' boson. Following these, there was a development by Weinberg and by Callan, Coleman, Wess and Zumino of chiral Lagrangians as generating functions for all the soft pion theorems.  This work was incorporated into chiral perturbation theory, an important non-perturbative calculational tool in the
standard model.

\section{Deep Inelastic Scattering Sum Rules}

{\bf Featured Paper;}  ``Sum Rules Giving Tests of Local Commutation Relations in High-Energy Neutrino Reactions'', by Stephen L. Adler, CERN and Lyman Laboratory,
Harvard University, Phys. Rev. {\bf 143}, 1144 (1966).

Let me turn next to deep inelastic scattering sum rules. As I noted, combining the forward lepton theorem with the $g_A$ sum rule led to  a sum rule for deep inelastic neutrino-nucleon scattering with a forward lepton. When I was at CERN in 1965 Gell-Mann urged me to try and generalize it to non-forward scattering. When I was sitting at a breakfast table at the Lake Garda home of my wife's cousin, while they were off on an expedition, I finally got all the kinematics to come together,  and what came out is this sum rule, in a somewhat different notation from the one I originally used,
\begin{align}\label{deep}
2=&\int_0^\infty d\nu [W_2^{\bar\nu p}(q^2,\nu)-W_2^{\nu p}(q^2,\nu)]~~~,\cr
q=&k_\nu-k_\ell~~~,~~~\nu=E_\nu-E_\ell~~~.\cr
\end{align}

Notice that after you've integrated over $\nu$, the result is  independent of $q^2$.  This means that if the contributions to the integrand were all proportional to  a form factor squared, which is what you get from all the low-lying resonances, the sum rule could not be saturated.  A corollary of the sum rule of Eq. \eqref{deep} is that in
the limit as $E_\nu \to \infty$,
\begin{equation}\label{limit}
\lim_{E_\nu \to \infty} \left[\frac{d \sigma^{\bar \nu p}}{dq^2}-\frac{d \sigma^{ \nu p}}{dq^2}\right]=\frac{G_F^2}{\pi}~~~,
\end{equation}
is simply a constant proportional to the square of the Fermi constant $G_F$, independent of $q^2$.

{\bf Featured Paper:} ``Inequality for Electron and Muon Scattering from Nucleons'', by J. D. Bjorken, Stanford Linear Accelerator Center, Phys. Rev. Lett. {\bf 16}, 408 (1966).

Neutrino experiments are hard to do, but Bjorken used an isospin rotation to convert Eq. \eqref{limit} to something that could be tested in electron scattering
experiments at SLAC,
\begin{equation}\label{electron}
\lim_{E_\nu \to \infty} \frac{d(\sigma_p+\sigma_n)}{dq^2}>\frac{2\pi\alpha^2}{q^4}~~~.
\end{equation}
 The factor $q^4$ is just a virtual photon propagator, so you see again that when you multiply back by $q^4$ there's a hard kernel inside.

{\bf Featured Paper:}  ``Asymptotic Sum Rules at Infinite Momentum'', by J. D. Bjorken, Stanford Linear Accelerator Center, Phys. Rev. {\bf 179}, 1547 (1969).

The origin of the hard kernel was  explained by a famous paper of Bjorken, where he introduced the ``scaling'' idea that $\nu W_2(q^2,\nu) \to F_2(x)$ with $x$ the
scaling variable defined by $x=q^2/(2 M_N \nu)$.  Applying this to the deep inelastic scattering sum rule,  you see that when you do the integral over $\nu$ the $q^2$
dependence scales out, leaving  a hard kernel  independent of $q^2$.

\section{Anomalies and PCAC}

{\bf Featured Paper:} ``Axial-Vector Vertex in Spinor Electrodynamics'', by Stephen L. Adler, Institute for Advanced Study, Phys. Rev.  {\bf 177}, 2426 (1969).

{\bf Featured Paper:}  ``A PCAC Puzzle:  $\pi^0 \to  \gamma \gamma$ in the $\sigma$-Model'', by J. S. Bell and R. Jackiw,  CERN, Nuovo Cimento A {\bf 60}, 47 (1969).

We now come to anomalies. This is a big subject. The combined wisdom of my paper and the paper of Bell and Jackiw  led to the following formula, which is given in an appendix of my paper,
\begin{align}\label{pizero}
\partial^\mu{\cal F}_{3\mu}^5=&\frac{f_\pi}{\surd 2}\phi_{\pi^0}+S\frac{\alpha}{4\pi}F^{\xi\sigma}F^{\tau \rho}\epsilon_{\xi\sigma\tau\rho}~~~,\cr
S=&\sum_jg_jQ_j^2~~~.\cr
\end{align}
 Eq. \eqref{pizero} states that if you want to do PCAC in the presence of electromagnetism, you start with the original PCAC piece, but for the neutral pion case you have to add a term  proportional to the anomaly.  The anomaly term has a well-defined coefficient $S$, which is the sum of axial-vector couplings times hadronic constituent charges squared, and  this coefficient multiplies the usual Abelian anomaly.   If you apply this formula to the calculation of $\pi^0 \to \gamma \gamma$ decay,  the Sutherland-Veltman theorem implies that the left-hand side vanishes for kinematic reasons . So without the anomaly term you would conclude that $\pi^0 \to \gamma \gamma$ vanishes, but it doesn't vanish -- it is the dominant decay mode.  With the anomaly-corrected PCAC formula of Eq. \eqref{pizero}, you find that the decay amplitude is entirely determined by the triangle anomaly, with a coefficient $S$ that counts quark charges. Comparing with experiment, one finds that the $\pi^0 \to \gamma \gamma$
 matrix element for charge $2/3, ~ -1/3$  quarks is a factor of 3 too small,  whereas with $SU(3)$ ``colored'' quarks this factor of 3 is made up, and one
 gets the correct decay rate.

 {\bf Featured Paper:}  ``Absence of Higher-Order Corrections in the Anomalous Axial-Vector Divergence Equation'', by Stephen L. Adler and William A. Bardeen, Institute for
 Advanced Study, Phys. Rev. {\bf 182}, 1517 (1969).

 In order for this counting of quarks to make sense you have to know that higher order strong interaction corrections don't mess the counting up, and that's where the paper I wrote with Bill Bardeen came in.  We showed that the anomaly coefficient  $S$  has no higher loop contributions, so it  is not renormalized by strong radiative corrections, and does count quarks. In the same year as we wrote this paper, Bardeen extended the Abelian anomaly to non-Abelian theories. Later on, anomaly non-renormalization was
 related to the fact that the general anomaly is topological in nature, with a coefficient that is an integer multiple of the Abelian anomaly coefficient.

 That's all I can say about anomalies in a minute; there are entire chapters in the standard field theory texts, and monographs as well, devoted to anomalies.

\section{Historical Significance}
 Let me now survey, very briefly, the historical significance of all this.

 \begin{itemize}

\item {First of all, the fact that the current algebra--PCAC program gave a lot of results that agree with the experiment was an indication that quantum field theory was the right direction for understanding the strong interactions, and not the  reciprocal bootstrap. When I came into physics and attended conferences, every conference would have a Mandelstam diagram on the board, and the lore, the mantra was that hadronic physics would be explained by a reciprocal bootstrap. After the successes of  PCAC and current algebra, that idea faded.}

\item {It became clear that field theory really had a future despite the fact that the strong coupling constant is too big to do perturbation theory. This was an indirect approach -- you took certain general structural properties of what was hoped to be the strong interaction  Lagrangian, derived relations from them which were found to be true, which hinted that underneath there was a field theory with a Lagrangian possessing the assumed structural properties.}

\item {PCAC showed that the strong interactions have an approximate chiral invariance, which we now understand to be due to the fact that the light quark masses are very small compared to the QCD scale.}

\item {The success of the Gell-Mann current algebra  indicated that there's a non-abelian or Yang-Mills gauge structure of the electroweak sector.}

\item {The deep inelastic sum rules, with a hard kernel and scaling, led to the reality of quarks.}

\item{The anomaly calculation of $\pi^0 \to \gamma \gamma$  led Bardeen, Fritzsch and Gell-Mann to introduce a color tripling of quarks.}

\item{Cancellation of anomalies between different fermion species, a concept introduced in my paper, played a role later on in understanding the family structure of
the standard model.}

\item{Also later on, topology entered particle physics through the structure of the anomaly.}

\item{So each of these little bits of theory, these nice theoretical formulas, had important implications as hints of what the underlying strong interaction and electroweak structure should be. For an excellent book-length discussion, see Tian Yu Cao, ``From Current Algebra to Quantum Chromodynamics: A Case for Structural Realism'', Cambridge University Press, 2010.}

\end{itemize}

\section{Updates and Revivals}

I finally come to some updates and revivals.

\subsection{Updates on the Adler-Weisberger and $\pi \pi$ sum rules,  the Goldberger-Treiman relation, and the Adler sum rule}

{\bf Featured Paper:}  ``Evaluation of the axial-vector commutator sum rule for pion-pion scattering'', by Stephen L. Adler, Institute for Advanced Study, and
F. J. Yndur\'ain, Departamento de F\'isica Te\'orica, C-XI, Universidad Aut\'onoma de Madrid, Phys. Rev.  D {\bf 75}, 116002 (2007).

 About a decade ago I learned that Paco Yndur\'ain -- who unfortunately died not long after we did our work -- had accurate pion-pion scattering phase shifts attained by careful dispersion relation analysis. So I wrote to him and asked whether we can go back and look at the pion-pion scattering sum rule. That led to a paper in which this was done, and led to an updated evaluation of both the Adler-Weisberger and $\pi \pi$  scattering sum rules, and of the Goldberger-Treiman relation. These two sum rules were  shown to be good to  better than six percent with current data,  and the Goldberger-Treiman relation to be good to around 2.3 percent.

{\bf Featured Paper:} ``Adler sum rule'', by Stephen L. Adler, Scholarpedia, 4(6):8653 (2009).

The other thing I got into recently was writing an article for the Scholarpedia  about the deep inelastic sum rules. In it I updated the derivation to include the full three family Cabibbo-Kobayashi-Maskawa matrix for weak mixing.   This eliminates the Cabibbo angle dependence that appeared in the sum rule in my original paper;  when you put in the full CKM mixing matrix these go away and the sum rule total is just 2, as given above.

\subsection{The Axion, and composite Higgs}

 Finally, revivals. The most prominent revival of the weak pion ideas,  and soft pion theorems, is probably the axion. The axion is a very light pseudo-Nambu-Goldstone boson, and  so is  a neutral  pion analog that arises, as shown by Weinberg and Wilczek, from spontaneous breaking of the Peccei-Quinn symmetry, which was introduced to solve the  strong CP problem. Now that people so far have not found WIMPS -- weak interacting massive particles -- as dark matter candidates, a lot of prominence is being given to looking for axions or similar very light pseudo-Nambu-Goldstone bosons as a light dark matter candidate. Another revival that's faded a bit but I think may well come back is the idea that the Higgs boson itself may be a pseudo-Nambu-Goldstone boson.   This was introduced by Susskind and Weinberg with the catchy name Susskind gave it, ``Technicolor'', with more recent  ``little Higgs''  versions of the original idea. And now that people haven't found supersymmetry as an explanation for the particle hierarchy, I suspect that there will again be more activity focused on the idea that the Higgs boson may be a composite. Thus the foundational ideas I talked about are currently active in a different guise.

\end{document}